\documentclass[preprint,aps,superscriptaddress]{revtex4}
\usepackage{amssymb}
\usepackage{amsmath}
\usepackage{graphicx}
\usepackage{color}

\newcommand\beq{ \begin{eqnarray} }
\newcommand\eeq{ \end{eqnarray} }

\begin{document}

\title{Towards a Holographic Model of D-Wave Superconductors}
\author{Jiunn-Wei Chen\thanks{%
E-mail: jwc@phys.ntu.edu.tw}}
\affiliation{Department of Physics and Center for Theoretical Sciences, National Taiwan
University, Taipei 10617, Taiwan}
\author{Ying-Jer Kao\thanks{%
E-mail: yjkao@phys.ntu.edu.tw}}
\affiliation{Department of Physics and Center for Theoretical Sciences, National Taiwan
University, Taipei 10617, Taiwan}
\author{Debaprasad Maity\thanks{%
E-mail: debu.imsc@gmail.com}}
\affiliation{Department of Physics and Center for Theoretical Sciences, National Taiwan
University, Taipei 10617, Taiwan}
\affiliation{Leung Center for Cosmology and Particle Astrophysics\\
National Taiwan University, Taipei 106, Taiwan}
\author{Wen-Yu Wen\thanks{%
E-mail: steve.wen@gmail.com}}
\affiliation{Department of Physics and Center for Theoretical Sciences, National Taiwan
University, Taipei 10617, Taiwan}
\affiliation{Leung Center for Cosmology and Particle Astrophysics\\
National Taiwan University, Taipei 106, Taiwan}
\author{Chen-Pin Yeh\thanks{%
E-mail: chenpinyeh@gmail.com}}
\affiliation{Department of Physics and Center for Theoretical Sciences, National Taiwan
University, Taipei 10617, Taiwan}

\begin{abstract}
The holographic model for S-wave high $T_{c}$ superconductors developed by
Hartnoll, Herzog and Horowitz is generalised to describe D-wave
superconductors. The 3+1 dimensional gravitational theory consists of a
symmetric, traceless second-rank tensor field and a $U(1)$ gauge field in
the background of the AdS black hole. Below $T_{c}$ the tensor field which
carries the $U(1)$ charge undergoes the Higgs mechanism and breaks the $U(1)$
symmetry of the boundary theory spontaneously. The phase transition
characterised by the D-wave condensate is second order with the mean field
critical exponent $\beta =1/2$. As expected, the AC conductivity is
isotropic below $T_{c}$ and the system becomes superconducting in the DC
limit but has no hard gap.
\end{abstract}

\maketitle

\begin{flushright}
\end{flushright}

%\date{\today}

\section{Introduction}

%The high temperature superconductors (HTCSC) are found to have layer
%structures and on each layer, the gap of the fermionic excitation near the
%fermi surface has the D-wave structure. (* Is this statement correct and
%general? How about the iron based SC? *) The underlying mechanism is not
%well understood. It is speculated that there might be pairing between
%electrons due to strong couplings. (* More about this. *) The strong
%coupling, on one hand, could naturally give higher critical temperature $%%
%(T_{c})$ compared with a weakly coupled system. The analyses, on the other
%hand, are more difficult. In general, non-perturbative approaches, such as
%lattice calculations, are required for the analyses. However, except in some
%special cases, the fermion sign problem emerges to prevent us from getting
%reliable results using Monte Carlo simulations. Given the situation,
%alternative algorithms and approaches are valuable to tackle the problem.

One of the unsolved mysteries in modern condensed matter physics is the
mechanism of the high temperature superconducting (HTSC) cuprates \cite%
{Lee:2006}. These materials are layered compounds with copper-oxygen planes
and are doped Mott insulators with strong electronic correlations. The
pairing symmetry is unconventional and there is a strong experimental
evidence showing that it is D-wave\cite{Tsuei:2000}. It is speculated that
the pairing between electrons is mediated via strong anti-ferromagnetic spin
fluctuations in the system. A prominent strong coupling theory is proposed
by Anderson, called the resonant valence bond (RVB) theory, which describes
liquid state with spin-singlets. Upon hole doping, the N\'{e}el order is
destroyed and give rise to superconductivity\cite{Anderson}. Several gauge
theories have been proposed to formulate the RVB physics, by enforcing the
double occupation constraint in the strong coupling limit\cite{gaugetheories}%
. The problem is difficult due to the strong-coupling nature of the theory.
Although significant progress has been made in the past few years,
alternative approaches may be valuable to tackle the problem.

One alternative approach is the holographic correspondence between a
gravitational theory and a quantum field theory, which first emerged under
AdS/CFT correspondence
\cite{Maldacena:1997re,Gubser:1998bc,Witten:1998qj}. This method has
provided a useful and complimentary framework to describe strong interaction
systems without a sign problem (see e.g. \cite%
{Policastro:2001yc,Herzog:2006gh,Liu:2006ug,Gubser:2006bz,Herzog:2007ij,Hartnoll:2007ih,Hartnoll:2007ip,Hartnoll:2008hs}%
). In the original top-down approach, both the gravity side and the field
theory side of the theories are precisely known. Later applications assume
that the correspondence exists among different pair of theories and try to
make predictions from one side of the correspondence. More specifically, in
this bottom-up approach, usually the gravity side of the theory is
explicitly constructed with the desired symmetries, then physical
observables (matrix elements) of the field theory side are predicted through
the above mentioned correspondence.%
%Interesting results in quark-gluon interactions and
%condensed matter systems near quantum criticality are obtained \cite{xxx}.

Recently, a gravitational model of hairy black holes \cite%
{Gubser:2005ih,Gubser:2008px} have been used to model S-wave HTSC \cite%
{Hartnoll:2008vx,Hartnoll:2008kx,Horowitz:2009ij,Gubser:2008zu}. In those
class of models the Abelian symmetry of a complex scalar field is
spontaneously broken (i.e. the Higgs mechanism) below some critical
temperature. The Meissner effect was soon observed by including magnetic
field in the background\cite{Nakano:2008xc,Albash:2008eh}. The effect of
superconducting condensate on the holographic fermi surface has been studied
by calculating fermionic spectral function \cite%
{Chen:2009pt,Faulkner:2009am,Gubser:2009dt}.  Interestingly, the properties
of spectral function appeared to have similar behaviour to that found in the
angle resolved photo-emission experiment. Motivated by all of these s-wave
studies, holographic dual to P-wave superconductor has been proposed by
coupling a $SU(2)$ Yang-Mills field to the black hole, where a vector hair
develops in the superconducting phase\cite%
{Gubser:2008wv,Roberts:2008ns,Martin:2009plb,Pallab}. Behaviour of fermionic
spectral function has also been studied in those p-wave superconducting
background \cite{gubser}. So far, the bottom-up construction of 
holographic superconductor assuming the existence of 
gauge/gravity duality has been
discussed. However, in the string theory framework people
also have studied top-down approach considering various D-brane
configurations in the AdS black hole background \cite{Martin2}.

In this work, we try to construct a minimal gravitational model that models
D-wave HTSC. We replace the complex scalar field in \cite{Hartnoll:2008vx}
by a tensor field whose condensate breaks the symmetry spontaneously below $%
T_{c}$ and the condensate becomes zero and the symmetry is restored above $%
T_{c}$. The critical exponent $\beta $ gives the mean field value $1/2$. The
real part of the conductivity computed from  linear response has a delta
function at zero frequency which corresponds to static superconductivity
below $T_{c}$. Above $T_{c}$, the delta function disappears as expected and
the conductivity becomes constant in frequency.\textbf{\ }It is expected
that there is no \textquotedblleft hard gap\textquotedblright\ in the real
part of the conductivity and the conductivity should be isotropic even
though the condensate is not (for a model calculation, see \cite{iso-sigma}%
). Both features are seen in our results.

\section{A Holographic Model for D-Wave HTSC}

Our goal is to consider a minimal (3+1 dimensional) holographic model that
gives rise to (2+1 dimensional) D-wave superconductivity. The construction
will be similar to that of the S-wave case \cite{Hartnoll:2008vx} with a
spontaneous local $U(1)$ symmetry breaking in the bulk leading to a
spontaneous breaking of global $U(1)$ symmetry at the boundary. Thus,
strictly speaking, the boundary theory is a super-fluid. One can still study
the current-current correlator which could be interpreted as the
conductivity.

To have a D-wave condensate at the boundary, we introduce a charged tensor
field in our dual gravity theory. Assuming the D-wave condensate originating
from electron-electron pairing, the D-wave nature gives a symmetric wave
function for the pair. So, wave function for this electron pair has to be a
spin singlet such that its total wave function is anti-symmetric. A $3\times
3$ symmetric traceless tensor has $5$ components which can be used to
describe a D-wave state. We will promote this symmetric traceless tensor
field to include time components and denote the field as $B_{\mu \nu }$ ($%
\mu ,\nu =0,1,2,3$), i.e. $B_{\mu \nu }=B_{\nu \mu }$ and $B_{\mu }^{\mu }=0$%
. However, it is important to note that the interacting higher
spin fields, in general, require to satisfy additional constraints
in addition to the equations of motion to remove the unphysical
degrees of freedom, see the discussion in
\cite{Pauli,Vasiliev,Deser} for example. Observing that there is
no available consistent model for our purpose, we
would like to propose a truncated model which has sufficient
ingredients to catch some features of D-wave superconductor. It
would be an important but difficult task to construct a complete
theory which we would like to postpone for the future in order to
attain our simple goal through the present exercise.

%\textbf{Thus, }$B_{\mu \nu }$\textbf{\ has }$9$\textbf{\
%components. We do not try to eliminate some of the components by making }$%
%B_{\mu \nu }$\textbf{\ a gauge field or imposing constraints. }

\textbf{\ }The desired complete action including gravity, $U(1)$ gauge
field, tensor field and other auxiliary fields, may take the following form%
\cite{comment}:
\begin{eqnarray}
S &=&\frac{1}{2\kappa ^{2}}\int d^{4}x\sqrt{-g}\left\{ \left( R+\frac{6}{%
L^{2}}\right) +\mathcal{L}_{m}+\mathcal{L}_{a}\right\} ,  \notag \\
\mathcal{L}_{m} &=&-\frac{L^{2}}{q^{2}}\left[ (D_{\mu }B_{\nu \gamma
})^{\ast }D^{\mu }B^{\nu \gamma }+m^{2}B_{\mu \nu }{}^{\ast }B^{\mu \nu }+%
\frac{1}{4}F_{\mu \nu }F^{\mu \nu }\right] ,  \label{Lagrangian}
\end{eqnarray}%
where $R$ is the Ricci scalar, the $6/L^{2}$ term gives a negative
cosmological constant and $L$ is the AdS radius which will be set
to unity in the units that we use. $\kappa ^{2}=8\pi G_{N}$ is the
gravitational coupling. $D_{\mu }$ is the covariant derivative in
the black hole background ($D_{\mu }=\partial _{\mu }+iA_{\mu }$
in flat space), and $q$ and $m^{2}$ are the charge and mass
squared of $B_{\mu \nu }$, respectively. The terms associated with
auxiliary fields are included in $\mathcal{L}_{a}$. This
Lagrangian is the same as that appears in \cite{Pauli} in the 
flat space limit. However in the subsequent studies after \cite{Pauli}
it appeared that construction of higher spin 
field Lagrangian coupled to gravity or $U(1)$ gauge in a 
gauge invariant way is non-trivial even in Minkowski space \cite{deser2}.
But later on in the context of gravitational interaction
\cite{fradkin} it has been shown that this task may
have a solution in AdS space.
Also there exists a recent attempt to construct gauge 
invariant $U(1)$ charged massive spin 2 particles
in AdS space at linear approximation \cite{zinoviev}.
In spite of all these studies fully consistent formulation of
interacting higher spin gauge fields is still lacking. 
Keeping this in mind that the action in (\ref{Lagrangian}) 
without $\mathcal{L}_{a}$ contains
spurious degrees of freedom that may cause the instability. We
will proceed by assuming that the constrains can eliminate the
instabilities but still allow the D-wave condensation.

%\textbf{Note that although it looks
%like the }$B_{t\nu }$\textbf{\ component has a wrong sign kinetic term,
%however, its mass term also has the same sign. Thus, its on-shell frequency
%does not become imaginary and hence this does not signal an instability.}
%}
$\mathcal{L}_{m}$ might look more familiar with the rescaling
$B_{\mu \nu }\rightarrow qB_{\mu \nu }$ and $A_{\mu }\rightarrow
qA_{\mu }$. Here we also concentrate on the \textquotedblleft
probe limit\textquotedblright\ \cite{Hartnoll:2008vx} where the
back-reaction to the background can be ignored. This limit is
exact when $q\rightarrow \infty $. In the probe limit,
$\mathcal{L}_{m}$ can be treated as a perturbation on
top of the 3+1 dimensional AdS black hole background:%
\begin{equation}
ds^{2}=-g(r)dt^{2}+\frac{dr^{2}}{g(r)}+r^{2}(dx^{2}+dy^{2}),  \label{metric}
\end{equation}%
where $g(r)=r^{2}-\frac{r_{0}^{3}}{r}$ and $r_{0}$ is the horizon size. The
Hawking temperature for this black hole $T=\frac{3r_{0}}{4\pi }$.

As in S-wave case \cite{Hartnoll:2008vx}, electric field can exist in the
bulk by the appropriate choice of boundary conditions. The charged tensor
field, which can be considered as charged particles, experiences a force
under the electric field, with positive(negative) charges
repelled(attracted) away from(toward) the black hole. One the other hand,
the black hole tries to pull all the charged particles in it. At lower $T$,
the black hole is smaller and the gravitational pull is weaker. Thus, the
positively charged particles have a bigger chance to stay outside the
horizon and form the condensate. At very large $T$, the gravitational force
from the large black hole is strong enough to pull all the charged particles
into the horizon such that there is no condensate. Thus, we have a phase
transition.

We are interested in describing the D-wave SC in the continuum such that
there is a condensate on the $x$-$y$ plane on the boundary with
translational invariance. Rotational symmetry is broken down to $Z(2)$ with
the condensate changing its sign under a $\pi /2$ rotation on the $x$-$y$
plane. To incorporate these features, we use an ansatz for the $B_{\mu \nu }$
and the gauge field $A_{\mu }$, i.e.,
\begin{equation}
B_{\mu \nu }=\text{diagonal}\left( 0,0,f(r),-f(r)\right) ,~~A=\phi (r)dt.
\end{equation}

%((((*There is a zero mode corresponding to rotation on the $x$-$y$ plane*)))).

After plugging in this ansatz, we have the equation of motion for $B$
\begin{equation}
r^{2}f^{\prime \prime }(r)+r\left[ r\frac{g^{\prime }(r)}{g(r)}-2\right]
f^{\prime }(r)+\left\{ \frac{r^{2}\phi ^{2}(r)}{g(r)^{2}}-\frac{\left[
m^{2}r+2g^{\prime }(r)\right] r}{g(r)}\right\} f(r)=0,  \label{EOM1}
\end{equation}%
and the corresponding Maxwell's equation is
\begin{equation}
r^{2}\phi ^{\prime \prime }(r)+2r\phi ^{\prime }(r)-\frac{4f^{2}(r)\phi (r)}{%
r^{2}g(r)}=0,  \label{Max}
\end{equation}%
where the $^{\prime }$ is the derivative with respect to $r$.

We would like to choose the solution such that $\phi (r)$ has the asymptotic
form
\begin{equation}
\phi (r)\rightarrow \mu +\frac{\rho }{r}  \label{Phi}
\end{equation}%
near the boundary $\left( r\rightarrow \infty \right) $, where $\mu $ is
interpreted as the chemical potential and $\rho $ as the charge density in
the boundary theory. Here, we will first assume this and then show that
indeed this can be satisfied later. Now, the asymptotic of Eq.\ref{EOM1}) has
the asymptotic form

\begin{equation}
r^{2}f^{\prime \prime }(r)-\left( m^{2}+4\right) f(r)\simeq 0
\end{equation}%
near the boundary, which yields
\begin{eqnarray}
&&f(r)\rightarrow f_{0}r^{\Delta _{+}}+f_{1}r^{\Delta _{-}},  \notag \\
&&\Delta _{\pm }=\frac{1\pm \sqrt{17+4m^{2}}}{2}.  \label{10}
\end{eqnarray}%
If we interpret $f_{0}$ as the source and $f_{1}$ as the vacuum expectation
value (VEV) of the operator that couples to $B$ at boundary theory, we need $%
m^{2}\geq -4$ (and $\Delta _{-}\leq 0$) such that the $f_{1}$ term is
constant or vanishing at the boundary. After setting the source $f_{0}=0$
and using $\Delta _{-}\leq 0$, Eq.(\ref{Max}) indeed gives the asymptotic
solution of Eq.(\ref{Phi}). Note that the $f_{0}r^{\Delta _{+}}$ term does
not impose a constraint on $m^{2}$ by requiring that the third term on the
LHS of Eq.(\ref{Max}) to be smaller than the other two terms since we have
imposed $f_{0}=0$. One way to see this is to do the integration of the
differential equations from the boundary, then $f(r)\rightarrow
f_{1}r^{\Delta _{-}}$, $\phi (r)\rightarrow \mu +\frac{\rho }{r}$ satisfy
the asymptotic behaviors of Eqs.(\ref{EOM1}) and (\ref{Max}). The order
parameter of the boundary theory can be read off from the asymptotic
behavior of $B$,
\begin{equation}
\langle \mathcal{O}_{ij}\rangle =\left(
\begin{array}{cc}
f_{1} & 0 \\
0 & -f_{1}%
\end{array}%
\right)   \label{VEV}
\end{equation}%
where $(i,j)$ are the indexes in the boundary coordinates $(x,y)$.

It is useful to note that the action and the equations of motion are
invariant under the scaling%
\begin{eqnarray}
\left( t,r,x,y\right)  &\rightarrow &\left( t/c,cr,x/c,y/c\right) ,  \notag
\\
\left( r_{0},T,g(r)\right)  &\rightarrow &\left( cr_{0},cT,c^{2}g(r)\right) ,
\notag \\
\left( f(r),\phi (r)\right)  &\rightarrow &\left( c^{2}f(r),c\phi (r)\right)
.
\end{eqnarray}%
Thus, we can always scale $\mu \rightarrow 1$. This also helps to keep track
of the scaling dimension for observables, e.g., the scaling dimensions for $%
\mu $, $\rho $, and $f_{1}$ are 1, 2, and $2-\Delta _{-}$, respectively.
\begin{figure}[tbp]
\begin{center}
\includegraphics[scale=0.9]{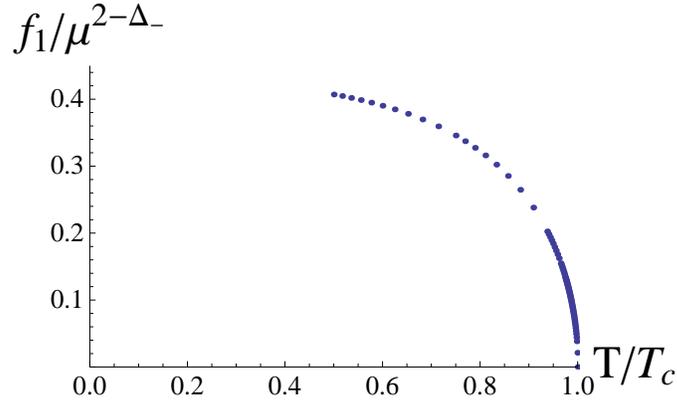}
\end{center}
\caption{(color online) The dimensionless D-wave condensate $f_{1}/\protect%
\mu ^{2-\Delta _{-}}$ shown as a function of $T/T_{c}$ for $m^{2}=-1/4$. The
condensate goes to zero at $T=T_{c}\propto \protect\mu $. The critical
exponent $f_{1}\rightarrow c\left( T_{c}-T\right) ^{\protect\beta }$ for $%
T_{c}-T\rightarrow 0^{+}$ is of the mean field value $\protect\beta =1/2$. }
\end{figure}

In Fig. 1, we show the numerical result between the dimensionless quantities
$f_{1}/\mu ^{2-\Delta _{-}}$ and $T/T_{c}$ for $m^{2}=-1/4$. It is a second
order phase transition. Numerically, the critical exponent $\beta $ defined
as $f_{1}\rightarrow c\left( T_{c}-T\right) ^{\beta }$ for $%
T_{c}-T\rightarrow 0^{+}$ is very close to the mean field value $\beta =1/2$%
. Below we show that only the values $\beta =1/2,3/2,5/2,\ldots $ satisfy
the equations of motion. Thus, without fine tuning, one would get the mean
field value $\beta =1/2$.

Now we present the derivation. The metric $g(r)$ is a smooth function of $%
\epsilon =T_{c}-T$, while $\phi $ and $f$ can be expanded as

\begin{eqnarray}
g(r,T) &=&a_{g}(r)+b_{g}(r)\epsilon +O\left( \epsilon ^{2}\right) ,  \notag
\\
\phi (r,T) &=&\epsilon ^{k}\left( a_{\phi }(r)+b_{\phi }(r)\epsilon +O\left(
\epsilon ^{2}\right) \right) ,  \notag \\
f(r,T) &=&\epsilon ^{n}\left( a_{f}(r)+b_{f}(r)\epsilon +O\left( \epsilon
^{2}\right) \right) .
\end{eqnarray}%
Since, Eq.(\ref{EOM1}) is a linear equation in $f$ and the pre-factors of $%
f^{\prime \prime }(r)$ and $f^{\prime }(r)$ are polynomials of $\epsilon $,
the pre-factor of $f(r)$ has to be a polynomial of $\epsilon $ as well in
order the satisfy the equation. This implies $k$ is an integer. At $T_{c}$
and at the boundary, $\phi $ gives the value of chemical potential which is
finite. This yields $k=0$.

Analogously, the Maxwell equation, Eq.(\ref{Max}), is linear in $\phi $. The
pre-factors of $\phi ^{\prime \prime }(r)$ and $\phi ^{\prime }(r)$ are
polynomials of $\epsilon $ and thus the pre-factor of $\phi (r)$ is required
to be a polynomial of $\epsilon $. This yields $2n$ to be an integer. We
also know that $n>0$ for a second order phase transition, and hence only $%
\beta =1/2,3/2,\ldots $are allowed.

\section{Conductivity}

In this section, we compute the conductivity of this D-wave HTSC by linear
response. The conductivity tensor $\sigma _{ij}$ can be defined through the
linear response relation
\begin{equation}
J_{i}=\sigma _{ij}E_{j},
\end{equation}%
where $i,j=1,2,$. $J$ and $E$ are the electric current and electric field,
respectively. Following the approach of \cite{Hartnoll:2008vx}, we perturb
the gauge field by $\delta A=e^{-i\omega t}A_{x}(r)dx$. To get a consistent
set of equations, we also need to perturbe $\delta B_{rx}=\delta
B_{xr}=ib_{rx}(r)e^{-i\omega t}$ and $\delta B_{tx}=\delta
B_{xt}=b_{tx}(r)e^{-i\omega t}$ respectively. The resulting equations of
motion are

\begin{equation}
gA_{x}^{\prime \prime }+g^{\prime }A_{x}^{\prime }+\left( \frac{\omega ^{2}}{%
g}-\frac{4f^{2}}{r^{4}}\right) A_{x}=0,  \label{Ax}
\end{equation}%
\begin{eqnarray}
&&gb\text{$_{rx}$}^{\prime \prime }+2g^{\prime }b\text{$_{rx}$}^{\prime }+%
\left[ \frac{g^{\prime \prime }}{2}-\text{ }\frac{g^{\prime }}{r}-5\text{ }%
\frac{g}{r^{2}}+r_{0}^{3}+\frac{(\omega -\phi )^{2}}{g}\right] b_{rx}  \notag
\\
&=&-\frac{2f}{r^{3}}A_{x}-\frac{(\omega -\phi )g^{\prime }}{g^{2}}b\text{$%
_{tx},$}  \label{B1}
\end{eqnarray}%
\begin{equation}
gb\text{$_{tx}$}^{\prime \prime }+\left[ -\frac{g^{\prime \prime }}{2}-\frac{%
g^{\prime }}{r}-\text{ }\frac{g}{r^{2}}+r_{0}^{3}+\frac{(\omega -\phi )^{2}}{%
g}\right] b\text{$_{tx}$}=(\omega -\phi )g^{\prime }b\text{$_{rx}.$}
\label{B1x}
\end{equation}

In principle, we can also add other $\delta B_{\mu \nu }$\ components in the
perturbation. However, those components do not couple to $\delta A_{\mu }$
to the quadratic order in the action. So, if we set the initial condition of
our system to be in the ground state before $\delta A_{\mu }$ perturbation
being turned on then those extra $B$ field perturbations will not be
produced. However, in the full stability analysis, those $\delta A_{\mu }$
independent perturbations are important. We will defer this stability
analysis to future study.

Eq.(\ref{Ax}) is very similar to the S-wave case and is decoupled from $%
\delta B_{\mu \nu }$. Near the boundary, we have%
\begin{equation}
r^{2}A_{x}^{\prime \prime }+2rA_{x}^{\prime }\simeq 0,
\end{equation}%
which yields the asymptotic form%
\begin{equation}
A_{x}\rightarrow A_{x,0}+\frac{A_{x,1}}{r},
\end{equation}%
where $A_{x,0}$ is the $x$-component gauge field at the boundary whose time
derivative gives $E_{x}$, and $A_{x,1}$ is the expectation value of the
current operator $J_{x}$. The ratio of $J_{x}$ and $E_{x}$ is the frequency
dependent conductivity
\begin{equation}
\sigma \left( \omega \right) \equiv \sigma _{xx}\left( \omega \right) =-%
\frac{iA_{x,1}}{\omega A_{x,0}}.
\end{equation}

\bigskip The fact that Eq.(\ref{Ax}) depends only on $A_{x}$ implies%
\begin{equation}
\sigma _{yx}\left( \omega \right) =0\text{.}
\end{equation}%
This is dictated by the reflection symmetry with respect to the $y=0$ plane.

We are now focusing on the case $m^{2}>-2$,
%(this is consistent with the requirement $m^{2}\leq 4$
%discussed above)
where the asymptotic forms of Eqs.(\ref{B1}) and (\ref{B1x}) are
particularly simple:%
\begin{eqnarray}
r^{2}b\text{$_{rx}$}^{\prime \prime }+4rb\text{$_{rx}$}^{\prime }-\left(
m^{2}+6\right) b_{rx} &\simeq &0\text{$,$}  \label{A1} \\
r^{2}b\text{$_{tx}$}^{\prime \prime }-\left( m^{2}+4\right) b\text{$_{tx}$}
&\simeq &0\text{$.$}
\end{eqnarray}%
These two equations can be solved with
\begin{eqnarray}
b\text{$_{rx}$} &\rightarrow &b\text{$_{rx,0}$}r^{\widetilde{\Delta }_{+}}+b%
\text{$_{rx,1}$}r^{\widetilde{\Delta }_{-}}  \notag \\
b\text{$_{tx}$} &\rightarrow &b\text{$_{tx,0}$}r^{\Delta _{+}}+b\text{$%
_{tx,1}$}r^{\Delta _{-}},  \label{A3}
\end{eqnarray}%
where $\widetilde{\Delta }_{\pm }=\frac{-3\pm \sqrt{33+4m^{2}}}{2}$ and $%
\Delta _{\pm }$ is defined in Eq.(\ref{10}). Here, we also identify $b_{rx,0}
$ and $b_{tx,0}$ as the source terms and $b_{rx,1}$ and $b_{tx,1}$ are the
normalizable fluctuations.%corresponding condensates.

Near the horizon, $g(r)=3r_{0}dr+\mathcal{O}(dr^{2})$ with $dr=r-r_{0}$ and $%
\phi (r)=\mathcal{O}(dr)$. The equations of motion become%
\begin{equation}
9dr^{2}A_{x}^{\prime \prime }+9drA_{x}^{\prime }+\frac{\omega ^{2}}{r_{0}^{2}%
}A_{x}=0,  \label{E1}
\end{equation}

\begin{equation}
9dr^{2}b\text{$_{rx}$}^{\prime \prime }+18drb\text{$_{rx}$}^{\prime }+\frac{%
\omega ^{2}}{r_{0}^{2}}b_{rx}=-\frac{\omega }{r_{0}^{2}dr}b\text{$_{tx}-%
\frac{6f(r_{0})}{r_{0}^{4}}dr$}A\text{$_{x},$}  \label{E2}
\end{equation}%
\begin{equation}
9dr^{2}b\text{$_{tx}$}^{\prime \prime }+\frac{\omega ^{2}}{r_{0}^{2}}b\text{$%
_{tx}$}=9\omega drb\text{$_{rx}.$}  \label{E3}
\end{equation}%
The solutions near the horizon are%
\begin{eqnarray}
A\text{$_{x}$} &\rightarrow &\overline{a}\text{$_{x,1}d$}r^{-i\frac{\omega }{%
3r_{0}}}+\overline{a}\text{$_{x,2}d$}r^{i\frac{\omega }{3r_{0}}},  \notag \\
b\text{$_{tx}$} &\rightarrow &3ir_{0}\left( \overline{b}\text{$_{rx,1}d$}%
r^{-i\frac{\omega }{3r_{0}}+1}-\overline{b}\text{$_{rx,2}d$}r^{-i\frac{%
\omega }{3r_{0}}}-\overline{b}\text{$_{rx,3}d$}r^{i\frac{\omega }{3r_{0}}+1}+%
\overline{b}\text{$_{rx,4}d$}r^{i\frac{\omega }{3r_{0}}}\right) ,  \notag \\
b\text{$_{rx}$} &\rightarrow &\overline{b}\text{$_{rx,1}d$}r^{-i\frac{\omega
}{3r_{0}}}+\overline{b}\text{$_{rx,2}d$}r^{-i\frac{\omega }{3r_{0}}-1}+%
\overline{b}\text{$_{rx,3}d$}r^{i\frac{\omega }{3r_{0}}}+\overline{b}\text{$%
_{rx,4}d$}r^{i\frac{\omega }{3r_{0}}-1}.
\end{eqnarray}

The ingoing wave boundary condition \cite{Liu:2009, Herzog:2003}, which sets
the wave falling into the horizon, demands $\overline{a}_{x,2}=\overline{b}%
_{rx,3}=\overline{b}_{rx,4}=0$. We further set the divergent term $\overline{%
b}_{rx,2}=0$ to keep the action finite. Now, $b_{rx,0(1)}$ and $b_{tx,0(1)}$
in the Eq.(\ref{A3}) are linear combinations of $\overline{a}_{x,1}$ and $%
\overline{b}_{rx,1}$. So we have both normalizable and non-normalizable
solutions for $b_{tx}$ and $b_{rx}$ perturbations.  The divergent source
terms $b_{rx,0}$ and $b_{tx,0}$ near the boundary can be cancelled by
counter terms \cite{Hartnoll:2008kx}.

\begin{figure}[tbp]
\begin{center}
\includegraphics[scale=0.7]{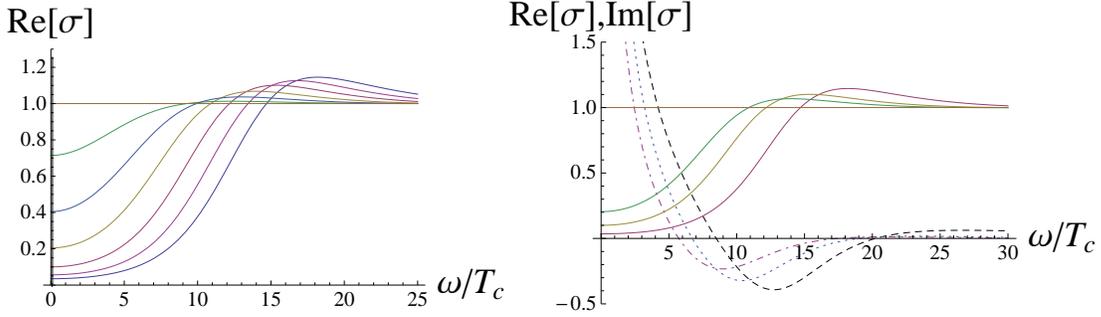}
\end{center}
\caption{(color online) The real (left plot) and imaginary (right plot) of
conductivity shown as a function for frequency $\protect\omega $ for
different temperatures. Above $T_{c}$, $Re[\protect\sigma \left( \protect%
\omega \right) ]=1$ while $Im[\protect\sigma \left( \protect\omega \right)
]=0$. Below $T_{c}$, $Re[\protect\sigma \left( \protect\omega \right) ]$ has
a $\protect\delta \left( \protect\omega \right) $ delta function whose
height decrease in $T$ and vanishes at $T_{c}$. The right most curve has the
lowest $T$, which implies the zero temperature gap $\protect\omega_g/T_c
\simeq 13$. (The construction in \protect\cite{Hartnoll:2008vx} for S-wave
gives the value $8$ for this gap.)}
\end{figure}

In Fig. 2, we plot the real and imaginary part of $\sigma \left( \omega
\right) $ for different $T$. The behaviours are similar to that of S-wave
HTSC. $Re[\sigma \left( \omega \right) ]$ has a delta function behaviour at $%
\omega =0$ corresponding to infinite DC conductivity when $T<T_{c}$. On the
othe hand for $T\geq T_{c}$, the delta function and $Im[\sigma \left( \omega
\right) ]$\ disappear and $Re[\sigma \left( \omega \right) ]$ becomes $%
\omega $ independent. There is no \textquotedblleft hard gap
\textquotedblright\ in our dual boundary superconducting system because $%
Re[\sigma \left( \omega \right) ]$ does not vanish even for arbitrary small $%
\omega $. One can also read off this soft gap from the plot, i.e. $\omega
_{g}/T_{c}\simeq 13$. It is larger than the one obtained in the construction
for S-wave\cite{Hartnoll:2008vx}, where $\omega _{g}/T_{c}\simeq 8$. This
may imply our D-wave pairing requires higher energy than the S-wave one.

Unlike the case for S-wave superconductor, the vanishing of the gap is
actually expected in the D-wave case. In the BCS-type theory (see, e.g. \cite%
{Dwave BCS}), the lowest dimensional D-wave operator for two fermion pairing
is $O_{ij}=\psi ^{T}\left( \overleftrightarrow{\partial }_{i}%
\overleftrightarrow{\partial }_{j}-\overleftrightarrow{\partial }^{2}\delta
_{ij}/2\right) \psi $, where $\overleftrightarrow{\partial }_{i}=%
\overrightarrow{\partial }_{i}-\overleftarrow{\partial }_{i}$\ is the
relative momentum between the two fermion which is invariant under Galilean
transformation. The leading order Lagrangian in a weakly interaction theory
is
\begin{equation}
\mathcal{L}=\mathcal{L}_{0}-c\left( O_{ij}^{\dagger }+J_{ij}^{\dagger
}\right) \left( O_{ij}+J_{ij}\right) +cJ_{ij}^{2},
\end{equation}%
where $\mathcal{L}_{0}$\ is the free Lagrangian, $c$\ is the coupling and $%
J_{ij}$\ is an external source. Under a Hubbard Stratanovich transformation,
the Lagrangian can be rewritten as
\begin{equation}
\mathcal{L}^{\prime }=\mathcal{L}_{0}+\left[ B_{ij}^{\dagger }\left(
O_{ij}+J_{ij}\right) +\left( O_{ij}^{\dagger }+J_{ij}^{\dagger }\right)
B_{ij}\right] +\frac{B_{ij}^{\dagger }B_{ij}}{c}+cJ_{ij}^{2},
\end{equation}%
where $B$\ and $B^{\ast }$\ are auxiliary fields. After integrating over the
auxiliary fields, $\mathcal{L}$\ is recovered from $\mathcal{L}^{\prime }$.
It is clear that $\left\langle O_{ij}\right\rangle $\ in $\mathcal{L}$\ is $%
\left\langle B_{ij}\right\rangle $\ in $\mathcal{L}^{\prime }$. The gap
equation of $\mathcal{L}^{\prime }$\ gives the dispersion relation
\begin{equation}
E_{k}=\sqrt{\left( \frac{k^{2}}{2m}-\mu \right) ^{2}+\left\vert
B_{ij}k_{i}k_{j}\right\vert ^{2}}.
\end{equation}%
The gap $\left\vert B_{ij}k_{i}k_{j}\right\vert \propto \left\vert
k_{x}^{2}-k_{y}^{2}\right\vert $\ vanishes at four nodes $%
k_{x}^{2}=k_{y}^{2} $. So, naturally gapless excitations can contribute to
conductivity. This makes the conductivity for a D-wave superconductor
gapless. In the S-wave case, however, the gap is isotropic and does not
vanish in any direction leading to a hard gap in conductivity.

If we change the gauge field perturbation to $\delta A=e^{-i\omega t}\left(
A_{x}(r)dx+A_{y}(r)dy\right) $, then there will be response from $\delta
B_{rx}$,$\delta B_{tx},\delta B_{ry}$ and $\delta B_{ty}$. $A(r)=A_{x}(r)%
\widehat{x}+A_{y}(r)\widehat{y}$ satisfies the same differential equation as
Eq.(\ref{Ax}):
\begin{equation}
g\mathbf{A}^{\prime \prime }+g^{\prime }\mathbf{A}^{\prime }+\left( \frac{%
\omega ^{2}}{g}-\frac{4f^{2}}{r^{4}}\right) \mathbf{A}=0.
\end{equation}%
This shows that the conductivity is isotropic:%
\begin{equation}
\sigma _{ij}\left( \omega \right) =\sigma \left( \omega \right) \delta _{ij}.
\end{equation}%
This might seem surprising at the first sight because the condensate is not
isotropic. However, this is a consequence of the symmetries that $\sigma
_{ij}$\ has in the D-wave case. In the linear response theory, $\sigma _{ij}$%
\ is a current-current correlator which can be schematically denoted as $%
\sigma _{ij}\sim \left\langle \Omega \left\vert \left[ J_{i},J_{j}\right]
\right\vert \Omega \right\rangle $, where the matrix element denotes an
ensemble average. Under a $\pi /2$\ rotation along the $z$-axis ($R$), $%
R^{-1}J_{i}R=\epsilon _{ij}J_{j}$, where $\epsilon _{ij}$\ is an
anti-symmetric tensor, and assuming the ensemble average is governed by
properties of the ground state which has the condensate structure of Eq.(\ref%
{VEV}), so $R\left\vert \Omega \right\rangle =-\left\vert \Omega
\right\rangle $. Then, $\sigma _{ij}\sim $\ $\left\langle \Omega \left\vert %
\left[ J_{i},J_{j}\right] \right\vert \Omega \right\rangle =\left\langle
\Omega \left\vert R^{-1}\left[ J_{i},J_{j}\right] R\right\vert \Omega
\right\rangle =\left\langle \Omega \left\vert \left[ \epsilon
_{ik}J_{k},\epsilon _{jl}J_{l}\right] \right\vert \Omega \right\rangle $.
This implies $\sigma _{xx}=\sigma _{yy}$\ and $\sigma _{xy}=-\sigma _{yx}$.
A similar analysis with parity operator with respect to the $x$-axis gives $%
\sigma _{xy}=\sigma _{yx}=0$. Thus, we have $\sigma _{ij}\propto \delta
_{ij} $. An explicit microscopic model calculation\ \cite{iso-sigma} also
yields an isotropic conductivity for a D-wave superconductor.

\section{Conclusion}

We have constructed a minimal holographic model for high $T_{c}$ D-wave
superconductors. We follow closely the work of Hartnoll, Herzog and Horowitz
on the S-wave case. The 3+1 dimensional gravitational theory consists a
symmetric, traceless second-rank tensor field and a $U(1)$ gauge field in
the background of the AdS black hole. Below $T_{c}$, the tensor field is
Higgsed to break the $U(1)$ symmetry at the boundary theory. The phase
transition characterised by the D-wave condensate is second order with the
mean field critical exponent $\beta =1/2$. As expected,\textbf{\ }the AC
conductivity is isotropic; below $T_{c}$, the system becomes superconducting
in the DC limit but has no hard gap.

\begin{acknowledgments}
We thank Sean Hartnoll, Pei-Ming Ho, Carlos Hoyos-Badajoz, Shamit Kachru,
Hsien-Chung Kao and Feng-Li Lin for useful discussions and Udit Raha for
careful reading of the manuscript. This work was supported in part by the
National Science Council and National Centre for Theoretical Sciences of
R.O.C. under grants NSC Grant Nos. 96-2112-M-002-019-MY3 (JWC) 97-2628-M-002 -011 -MY3
(YJK), 97-2112-M-002-015-MY3 (WYW).
\end{acknowledgments}

%%%%%%%%%%%%%%%%%%%%%%%%%%%%%%%%%%%%%%%%%%%%%%%%%%%%%%%%%%%%%%


\begin{thebibliography}{99}
\bibitem{comment} One can also include kinetic terms $\kappa _{2}(D^{\mu
}B_{\mu \nu })^{\ast }D_{\gamma }B^{\gamma \nu }$\ and $\kappa _{3}(D_{\mu
}B_{\nu \gamma })^{\ast }D^{\nu }B^{\mu \gamma }$\ and also potential terms
of the $B$\ field in Eq.(\ref{Lagrangian}). However, these terms will make
the analysis of conductivity more complicated since the perturbation of $%
A_{\mu }$\ does not decouple from the perturbation of $B_{\mu \nu }$\ unless
$\kappa _{3}=2$.
\end{thebibliography}
\end{document}